\documentclass[aps,pra,preprint,groupedaddress,superscriptaddress,showpacs]{revtex4-1}
\usepackage{CJK}
\usepackage{amsmath}
\usepackage{longtable}
\usepackage{mathrsfs}
\usepackage{multirow}
\usepackage{tabularx}
\usepackage{array}
\usepackage{graphicx}
\usepackage{bm}
\begin{document}
%\begin{CJK*}{GB}{}
%\preprint{}

\title{Reevaluation of the nuclear electric quadrupole moment for $^{87}$Sr by hyperfine structures and relativistic atomic theory}

\author{Benquan Lu}
\affiliation{Institute of Applied Physics and Computational Mathematics, 100088 Beijing, China}
\affiliation{National Time Service Center, 710600 Lintong, China}
\affiliation{The University of Chinese Academy of Sciences, 100088 Beijing, China}

\author{Tingxian Zhang}
\affiliation{Institute of Applied Physics and Computational Mathematics, 100088 Beijing, China}
\affiliation{Wuhan Institute of Physics and Mathematics, 430071 Wuhan, China}
\affiliation{The University of Chinese Academy of Sciences, 100088 Beijing, China}

\author{Hong Chang}
\affiliation{National Time Service Center, 710600 Lintong, China}
\affiliation{The University of Chinese Academy of Sciences, 100088 Beijing, China}

\author{Jiguang Li}
\email{li\_jiguang@iapcm.ac.cn}
\affiliation{Institute of Applied Physics and Computational Mathematics, 100088 Beijing, China}

\author{Yong Wu}
\affiliation{Institute of Applied Physics and Computational Mathematics, 100088 Beijing, China}

\author{Jianguo Wang}
\affiliation{Institute of Applied Physics and Computational Mathematics, 100088 Beijing, China}

\date{\today}

\begin{abstract}
The values of nuclear electric quadrupole moment are different by about 7\% for $^{87}$Sr nucleus between the recommended value [N. J. Stone, At. Data Nucl. Data Tables \textbf{111-112}, 1 (2016); P. Pyykk\"{o}, Mol. Phys. \textbf{116}, 1328 (2018)] and earlier results [e.g. A. M. M{\aa}tensson-Pendrill, J. Phys. B: At. Mol. Opt. Phys. \textbf{35}, 917 (2002); K. Z. Yu \textit{et al.}, Phys. Rev. A \textbf{70}, 012506 (2004)]. In this work, we reported a new value, $Q$($^{87}$Sr)~=~328(4)~mb, making use of our calculated electric field gradients produced by electrons at nucleus in combination with experimental values for hyperfine structures of the $5s5p~^3P_{1,2}$ states of the neutral Sr atom. In the framework of the multi-configuration Dirac-Hartree-Fock theory, the electron correlations were taken into account systematically so as to control the uncertainties of the electric field gradient at about 1\% level. The present result is different from the recommended value, but in excellent agreement with those by M{\aa}tensson-Pendrill and Yu~\textit{et al.}. We would recommend the present $Q$ value as a reference for $^{87}$Sr.
\end{abstract}

%\keywords{}

\maketitle
%\end{CJK*}
\section{Introduction}
The nuclear electric quadrupole moment $Q$ is an important parameter as it, together with the nuclear magnetic dipole moment $\mu_I$, can determine hyperfine structures in atoms and test nuclear models. In addition, the quadrupole moment is a unique and excellent tool to study nuclear deformation and shape coexistence~\cite{0034-4885-66-4-205, RevModPhys.83.1467}, especially for exotic nuclei in the vicinity of the proton and the neutron drip lines~\cite{Campbell2016127}.

The neutron-rich strontium isotopes are particularly good examples as their nuclei exhibit extremely deformed and spherical configurations for different isotopes~\cite{LIEVENS1991141, PhysRevC.93.014315}. This coexistence is connected with a rapid transition from spherical to deformed shapes, which is difficult to explain in the unified description of shape coexistence~\cite{PhysRevLett.116.022701}. The electric quadrupole moments $Q$ of the strontium isotopes are significant parameters for understanding these phenomena. Recently, the nuclear quadrupole moments of the neutron-rich $^{96,98}$Sr isotopes were measured by safe Coulomb excitation of radioactive beams at REX-ISOLDE, confirming the shape coexistence for the first time~\cite{PhysRevLett.116.022701}. For the $^{77,79,83,85,89,91,93,99}$Sr isotopes, the nuclear electric quadrupole moments were obtained from the ratio of the electric quadrupole hyperfine interaction constants $B$, $^{A}{\rm Q}/^{87}{\rm Q}~=~^{A} B/^{87} B$, by using the quadrupole moment of $^{87}$Sr as a reference~\cite{PhysRevC.41.2883, LIEVENS1991141}. However, there exist several values of electric quadrupole moment of the stable $^{87}$Sr nucleus, and the largest differences among them are about 18\%~\cite{ZPhys1963, PRA161371, 0953-4075-35-4-315, PRA70012506, PRA73062501}. These discrepancies arise mainly from calculations on the electric field gradients (EFGs) produced by electrons at the nucleus, since the error bars of measured hyperfine interaction constants $B$ are much less than 18\%. Therefore, it is worthwhile and essential to reevaluate the EFG values.

Main difficulty in determination of the EFGs with high accuracy results from complicated electron correlations in the atomic system. In this work, we calculated the EFGs of the $5s5p~^{1,3}P_1$ and $^3P_2$ states of the neutral Sr atom by using the multi-configuration Dirac-Hartree-Fock (MCDHF) method. The electron correlation effects on the EFGs were investigated in detail. In order to estimate the uncertainties of the EFGs, we also calculated the magnetic dipole hyperfine interaction constants $A$ for these three states, since in general the $A$ values are more sensitive to the electron correlation effects. Combined our calculated EFGs with measured electric quadrupole hyperfine interaction constants available, a new nuclear quadrupole moment of $^{87}$Sr was given with an uncertainty of 1\%.

\section{Theory}
In the framework of the MCDHF method~\cite{Grant2007, Fischer_2016}, an atomic state function (ASF) $\Psi (\Gamma PJM_J)$ is a linear combination of configuration state functions (CSFs) $\Phi_j (\gamma_j PJM_J)$ with the same parity $P$, total angular momentum $J$, and its component along $z$ direction $M_J$, that is,
\begin{equation}\label{eq:1}
\Psi (\Gamma PJM_J)=\sum_{j}^{N}c_j\Phi_j (\gamma_jPJM_J).
\end{equation}
Here, $c_j$ represents the mixing coefficient corresponding to the $j^{th}$ configuration state function, and $\gamma$ stands for the other quantum numbers which can define the atomic state uniquely. The configuration state functions $\Phi_j(\gamma_jPJM_J)$ are built from sums of products of one-electron Dirac orbitals
\begin{equation}\label{eq:2}
\phi(r,\theta ,\varphi)=\frac{1}{r}\binom{P(r)\chi_{\kappa m}(\theta ,\varphi)}{iQ(r)\chi_{-\kappa m}(\theta ,\varphi)},
\end{equation}
where $P(r)$ and $Q(r)$ are the radial functions and $\chi_{\kappa m}(\theta, \varphi)$ are two-component spherical spinors. The mixing coefficients $c_j$ and the radial functions are optimized simultaneously in the self-consistent field (SCF) procedure. The Breit interaction and the quantum electrodynamic (QED) corrections can be included in the relativistic configuration interaction (RCI) computation, in which only mixing coefficients are varied.

Hyperfine structures of atomic energy levels are caused by the interaction between the electrons and the electromagnetic multipole moments of the nucleus. The corresponding Hamiltonian can be expressed as a multipole expansion
\begin{equation}\label{eq:3}
H_{hfs}=\sum_{k\geq 1}\mathbf{T}^{(k)}\cdot \mathbf{M}^{(k)},
\end{equation}
where $\mathbf{T}^{(k)}$ and $\mathbf{M}^{(k)}$ are spherical tensor operators of rank $k$ in the electronic and nuclear space, respectively. The $k = 1$ term represents the magnetic dipole hyperfine interaction, and the $k = 2$ the electric quadrupole hyperfine interaction. The higher-order, for instance, the nuclear magnetic octupole hyperfine interactions are negligible~\cite{PhysRevA.77.012512, PhysRevA.87.012512}. Furthermore, the magnetic dipole and the electric quadrupole hyperfine interaction constants ($A$ and $B$) are defined by
\begin{equation}\label{eq:4}
A= \frac{\mu_I}{I}  \left [ \frac{1}{J(J+1)} \right ]^{1/2}   \left \langle P J \left \| \mathbf{T}^{(1)} \right \| {PJ}\right \rangle,
\end{equation}
and
\begin{equation}\label{eq:5}
B=2Q \left [ \frac{J(2J-1)}{(J+1)(2J+3)} \right ]^{1/2}\left \langle PJ \left \| \mathbf{T}^{(2)} \right \| {PJ}\right \rangle.
\end{equation}
$\mu_I$ and $Q$ in the equations above are the nuclear magnetic dipole and electric quadrupole moments, respectively. The electronic tensor operators  $\textbf{T}^{(1)}$ and $\textbf{T}^{(2)}$ are given by
\begin{equation}\label{eq:6}
\textbf{T}^{(1)} = \sum_{j}-i\alpha r_{j}^{-2}( \bm{\alpha}_j \cdot \mathbf{l}_j \mathbf{C}^{(1)}(j)),
\end{equation}
\begin{equation}\label{eq:7}
\textbf{T}^{(2)} = \sum_{j}-r_{j}^{-3}\mathbf{C}^{(2)}(j),
\end{equation}
where $i$ is the imaginary unit, $r_j$ is the radial coordinate of the $j^{th}$ electron, $\mathbf{l}$ is the orbital angular momentum operator, $\mathbf{C}^{(k)}$ is a spherical tensor of rank $k$, $\alpha$ is the fine structure constant and $\bm{\alpha}_j$ is the Dirac matrix. The summation is made over $N$ electrons in the atom.

According to Eq.~(\ref{eq:5}), the nuclear electric quadrupole moment $Q$ (in mb) can be extracted from the $B$ constant through~\cite{PhysRevA.64.052507, 0953-4075-49-11-115002}
\begin{equation}\label{eq:8}
Q = \frac{4.2559579}{\mathrm{EFG}} B,
\end{equation}
where $B$ is in the unit of MHz and EFG (in a.u.), the electric field gradient produced by electrons at nucleus, is defined as
\begin{equation}\label{eq:9}
{\rm EFG} =2\left [\frac{J(2J-1)}{(J+1)(2J+3)} \right ]^{1/2}\left \langle PJ \left \| \mathbf{T}^{(2)} \right \| {PJ}\right \rangle.
\end{equation}

\section{Computational Models}\label{sec:3}
In this work we adopted the active space approach to capture electron correlations~\cite{CPL48157, JCP892185}. According to the perturbation theory, electron correlations can be divided into the first- and higher-order correlations~\cite{MCDHF1997, PhysRevA2012}. The corresponding configuration space was generated by single (S) and double (D) excitations from the occupied orbitals in the reference configuration(s) to a set of unoccupied orbitals. At the beginning, the single reference (SR) configuration was used to consider the first-order electron correlation which is composed of the correlation between valence electrons (VV correlation), the correlation between valence and core electrons (CV correlation) and the correlation between core electrons (CC correlation). Afterwards, the dominant CSFs in the first-order correlation function were selected to form the multi-reference (MR) configuration set. The CSFs generated by SD excitations from the MR configuration set can account for the main higher-order electron correlations.

Our calculation was started in the Dirac-Hartree-Fock (DHF) approximation, in which the occupied orbitals in the reference configuration $1s^22s^22p^63s^23p^63d^{10}4s^24p^65s5p$, also called spectroscopic orbitals, were optimized. Electrons in the outermost $5s$ and $5p$ orbitals in the reference configuration were treated as the valence electrons and the others the core. The VV correlation was considered in the SCF procedure through the configuration space expanded by SD-excitation CSFs from the $5s5p$ valence shells. As presented in Table~\ref{tab:1}, the unoccupied orbitals were augmented 
layer by layer to make convergence of parameters under investigation, and only the added orbitals were variable each time for making the average energy of the $5s5p$ configuration minimum. To raise computational efficiency, the CSFs which do not interact with the reference configurations were removed~\cite{Jonsson2007597, MCDHF1997}.

Subsequently, we took into account the CV correlations between electrons in the core and the valence electrons. The configuration state functions generated by restricted SD excitations from the single reference configuration were added to the VV computation model. The restricted SD excitations mean that at most one occupied electron in the core sub-shell can be substituted to the partially occupied or the unoccupied orbitals. The expansion of the configuration space and the optimization of the unoccupied orbitals were in the same way as that used in the VV computational model. Additionally, in order to analyse which core electrons strongly interact with the valence electrons, we opened up the subshell in the core one by one down to $n = 3$. Each step was labeled with C$_{nl}$V in Table~\ref{tab:1} where $nl$ represents the latest opened core subshell. The orbital set obtained in the C$_{3s}$V model was used for the subsequent RCI calculation.

The CV correlations between the electrons in the $n = 1, 2$ core shells and the valence electrons were estimated in the RCI computations, which were labeled with C$_{2p}$V, C$_{2s}$V and C$_{1s}$V, respectively. Furthermore, the CC electron correlation related to the $n = 4$ shell was included as well in the RCI calculation. In this step, the CSFs were generated by substituting one or two orbitals from the $n = 4$ core shell to the first five layers of the unoccupied orbitals. This computational model was marked as CC4-5 in Table~\ref{tab:1}.

The MR-SD model was applied to estimate the higher-order electron correlation effects among the $n = 4, 5$ shells on atomic parameters concerned. As mentioned earlier, the multi-reference configurations set was formed by selecting the dominant CSFs in the CC4-5 model, i.e. those CSFs with mixing coefficients $c_j$ larger than 0.04. Therefore, the \{$4s^24p^64d5p$; $4s^24p^65s6p$; $4s^24p^65p6s$; $4s^24p^64d6p$\} configurations were added to the SR configuration set. The SD excitations were allowed from the MR configurations to the first five layers of the unoccupied orbitals, which was marked as MR-5. This calculation was performed with the RCI method. Finally, the Breit interaction and QED corrections were evaluated.

It should be noted that in different stages of our calculations part of/all of orbitals were fixed. As a result, more unoccupied orbitals were required and higher-order correlations should be taken into account to achieve expected accuracy for physical quantities under investigation. This method sacrifices computational efficiency to a large extent. In order to deal with this issue, nonorthogonal orbital basis based on the pair-correlation functions (PCFs), accounting for different types of correlation effects, would be a potential excellent approach in the near future~\cite{Verdebout2010,Verdebout2013,FroeseFischer2013}

For convenience, the reference configuration(s) and the numbers of configuration state functions (NCSF) for states belonging to the $5s5p$ configuration in the different computational models are presented in Table~\ref{tab:1}. In practice, the GRASP2K package~\cite{Jonsson20132197} was employed to perform calculations.

\begingroup
\squeezetable
\begin{table}[!ht]
\caption{\label{tab:1}The numbers of configuration state functions (NCSF) for states in the $5s5p$ configuration generated in various computational models.}
\begin{ruledtabular}
\begin{tabular}{llcccc}
                                             &                     &         &         &   NCSF  &       \\ \hline
Occupied orbitals                            & Unoccupied orbitals & Model   & $J$=0   &   $J$=1 & $J$=2 \\ \hline
\{$5s5p$\}                                   &                     &  DHF    &     1   &      2  &      1     \\
\{$5s5p$\}                                   &\{4d,4f\}            &  VV     &     3   &      8  &      9     \\
                                             &\{6s,6p,5d,5f,5g\}   &         &    14   &     38  &     44     \\
                                             &\{7s,7p,6d,6f,6g\}   &         &    33   &     90  &    105     \\
                                             &\{8s,8p,7d,7f,7g\}   &         &    60   &    164  &    192     \\
                                             &\{9s,9p,8d,8f,7g\}   &         &    90   &    245  &    285     \\
                                             &\{10s,10p,9d,8f,7g\} &         &   117   &    315  &    360     \\
                                             &\{11s,11p,10d,8f,7g\}&         &   148   &    395  &    445     \\
\\
\{$4p^65s5p$\}                               &\{4d,4f\}            &C$_{4p}$V&    46   &    126  &     97     \\
                                             & ...                 &         &         &         &            \\
                                             &\{11s,11p,10d,8f,7g\}&         &  2724   &  12027  &  11417     \\
\\
\{$4s^24p^65s5p$\}                           &\{4d,4f\}            &C$_{4s}$V&    62   &    166  &    127      \\
                                             & ...                 &         &         &         &             \\
                                             &\{11s,11p,10d,8f,7g\}&         &  3671   &  15420  &  14488     \\
\\
\{$3d^{10}4s^24p^65s5p$\}                    &\{4d,4f\}            &C$_{3d}$V&   119   &    318  &    238     \\
                                             & ...                 &         &         &         &          \\
                                             &\{11s,11p,10d,8f,7g\}&         &  7141   &  31924  &  31168   \\
\\
\{$3p^63d^{10}4s^24p^65s5p$\}                &\{4d,4f\}            &C$_{3p}$V&   162   &    436  &    326    \\
                                             & ...                 &         &         &         &          \\
                                             &\{11s,11p,10d,8f,7g\}&         &  9717   &  43556  &  42140     \\
\\
\{$3s^23p^63d^{10}4s^24p^65s5p$\}            &\{4d,4f\}            &C$_{3s}$V&   178   &    476  &    356    \\
                                             & ...                 &         &         &         &          \\
                                             &\{11s,11p,10d,8f,7g\}&         & 10664   &  46949  &  45211     \\[0.2cm]
\{$2p^63s^23p^63d^{10}4s^24p^65s5p$\}        &\{11s,11p,10d,8f,7g\}&C$_{2p}$V& 13240   &  58581  &  56183     \\[0.2cm]
\{$2s^22p^63s^23p^63d^{10}4s^24p^65s5p$\}    &\{11s,11p,10d,8f,7g\}&C$_{2s}$V& 14187   &  61974  &  59254     \\[0.2cm]
\{$1s^22s^22p^63s^23p^63d^{10}4s^24p^65s5p$\}&\{11s,11p,10d,8f,7g\}&C$_{1s}$V& 15134   &  65367  &  62325     \\[0.2cm]
$\bigcup$\{$4s^24p^65s5p$\}                  &\{9s,9p,8d,8f,7g\}   &CC4-5    & 17638   &  85131  &  80059     \\
$\bigcup$\{$4s^24p^64d5p$;                   &\{9s,9p,8d,8f,7g\}   &MR-5     & 64717   & 345481  & 522345     \\
$4s^24p^65s6p$;                              &                     &         &         &         &              \\
$4s^24p^65p6s$;                              &                     &         &         &         &              \\
$4s^24p^64d6p$\}                             &                     &         &         &         &              \\
\end{tabular}
\end{ruledtabular}
\end{table}
\endgroup

\section{Results and discussion}
\subsection{CV Correlations from different core orbitals}
\begin{figure}
 \includegraphics[scale=0.9]{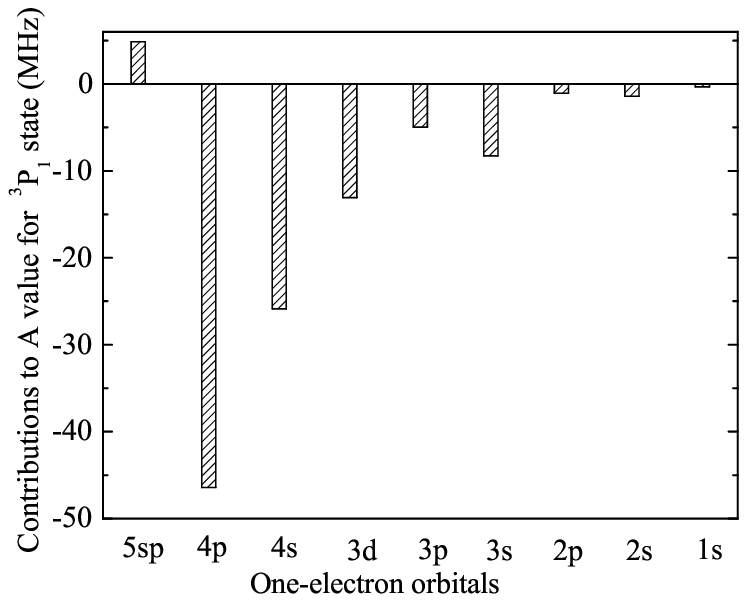}
 \includegraphics[scale=0.9]{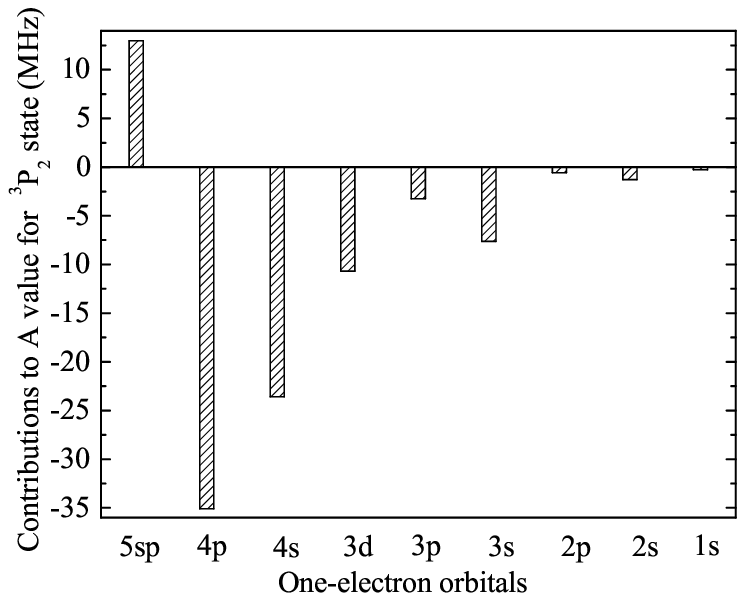}
 \includegraphics[scale=0.9]{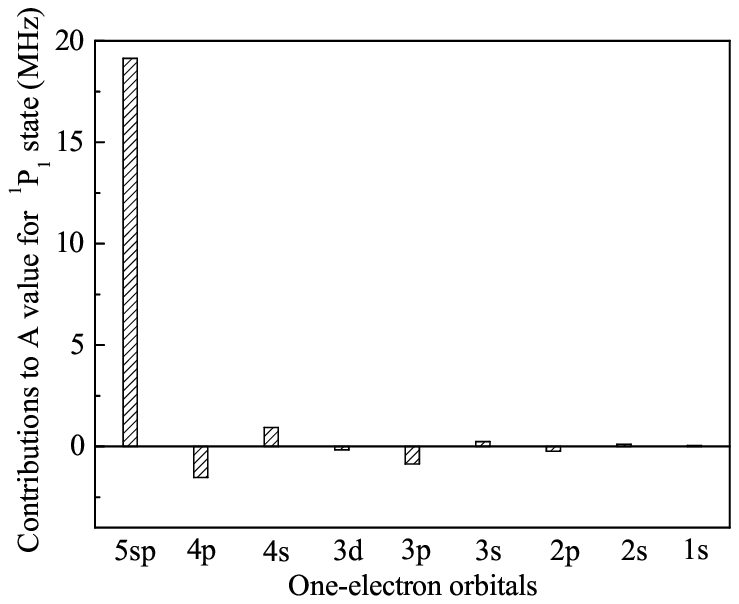}
 \caption{\label{fig:1}Contributions from occupied orbitals to the calculated values of $A$ for $^3P_1$, $^3P_2$ and $^1P_1$ states in $^{87}$Sr.}
\end{figure}

\begin{figure}
 \includegraphics[scale=0.9]{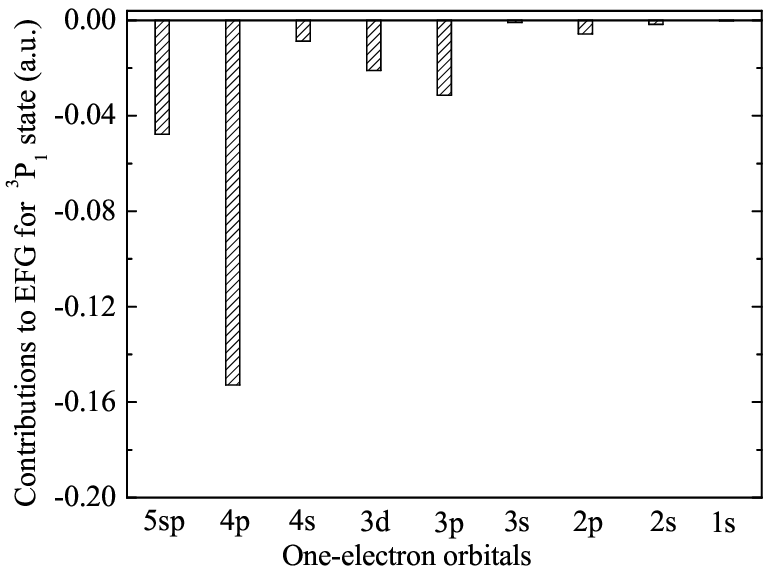}
 \includegraphics[scale=0.9]{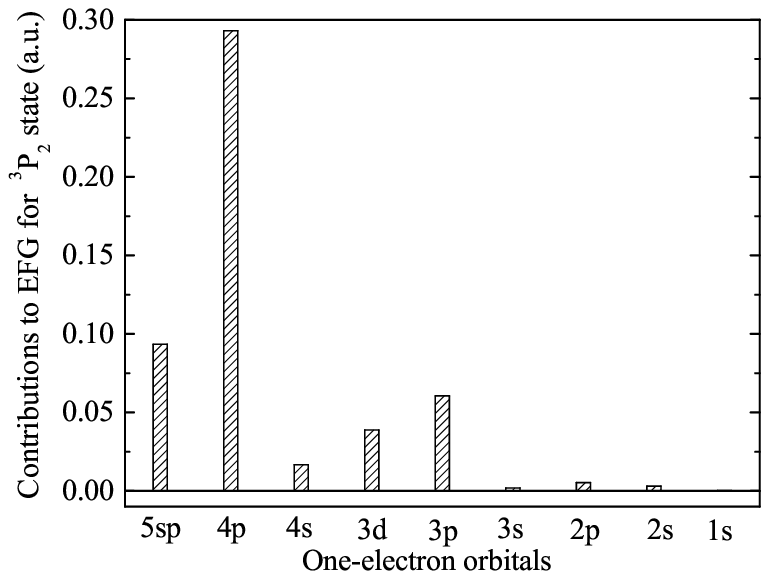}
 \includegraphics[scale=0.9]{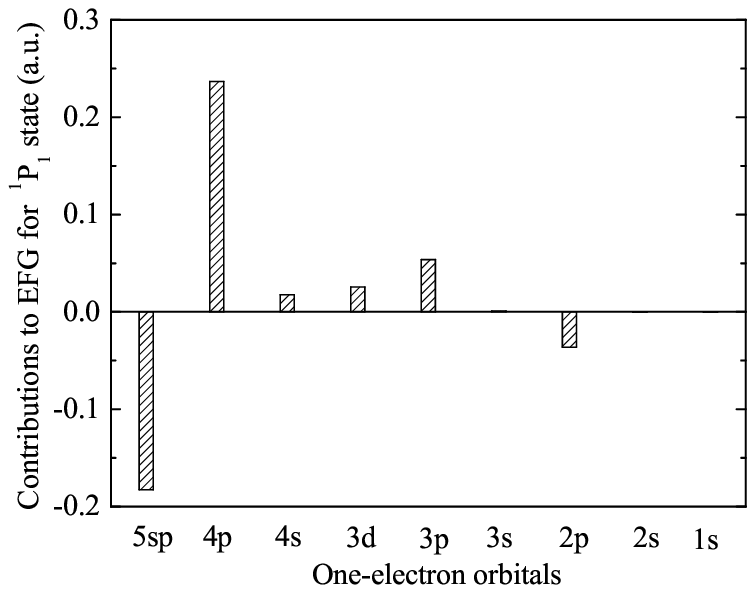}
 \caption{\label{fig:2}Contributions from occupied orbitals to the calculated values of EFG for $^3P_1$, $^3P_2$ and $^1P_1$ states in $^{87}$Sr.}
\end{figure}

As described above, we constructed several computational models to explore contributions from the different electron pairs to magnetic dipole hyperfine interaction constants $A$ and EFGs of $^3P_1$, $^3P_2$ and $^1P_1$ states. The results are shown in Fig.~\ref{fig:1} and~\ref{fig:2}. The influence of the CV correlation between the $4p$ core and the valence electrons on the $A$ constants and the EFGs, for example, was given by the difference in results between the C$_{4p}$V model and the VV model. From Fig.~\ref{fig:1}, it is clear that the magnetic dipole hyperfine interaction constants $A$ of $^3P_1$ and $^3P_2$ states are sensitive to the CV correlations, especially to the outermost core shell. Moreover, the contributions from the orbitals with $s$ and $d$ angular symmetrys are significant to the $A$ constant. For the EFGs, it was found in Fig.~\ref{fig:2} that the CV correlation effects related to the $4p$ orbital are more important than others. In addition, the orbitals with $p$ and $d$ angular symmetrys play key roles for the EFGs. We should also emphasized that the CV correlations between electrons in the $n = 1, 2$ core and the valence shells cannot be neglected for achieving high precision, although their contributions are quite small.

\begin{figure}
 \includegraphics[scale=0.9]{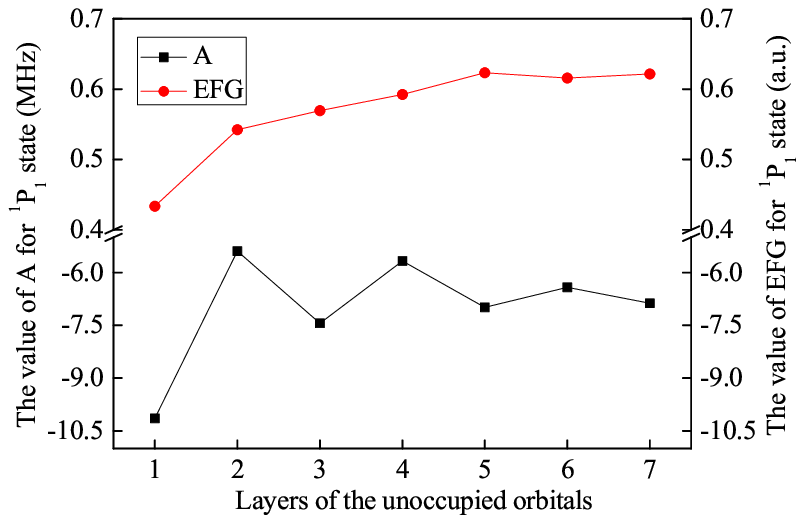}
 \includegraphics[scale=0.9]{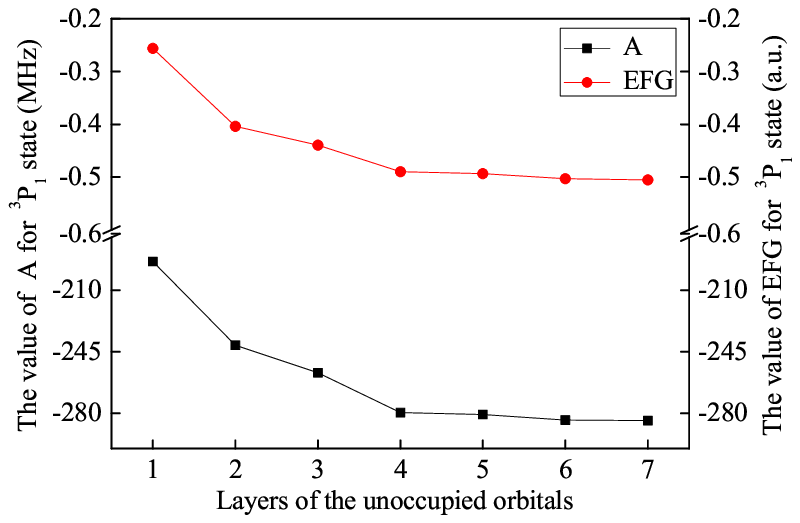}
 \includegraphics[scale=0.9]{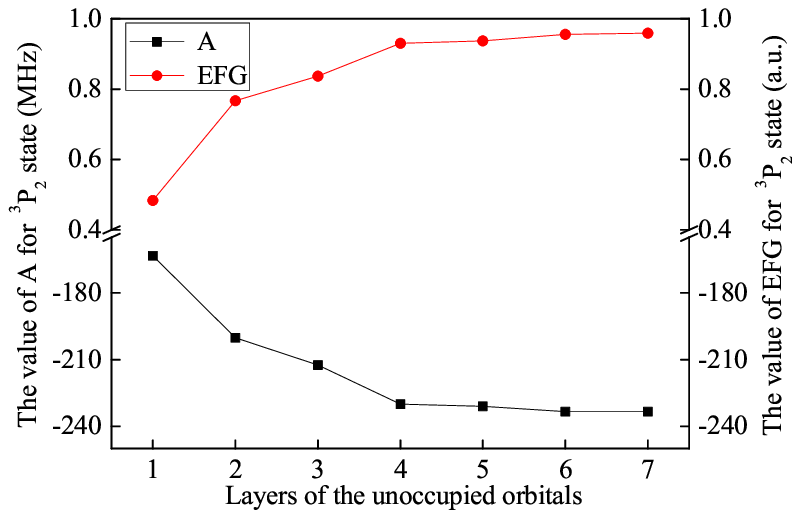}
 \caption{\label{fig:3}The values of $A$ constants and EFGs for $^1P_1$, $^3P_1$ and $^3P_2$ states as functions of layers of the unoccupied orbitals in C$_3$V model.}
\end{figure}

In order to check convergence of the parameters under investigation, we present in Fig.~\ref{fig:3} variations of the calculated $A$ constants and EFGs in C$_3$V model as functions of layers of the unoccupied orbitals.

\subsection{Uncertainty estimation for $A$ constants}
In Table~\ref{tab:2} we display the calculated magnetic dipole hyperfine interaction constants $A$ for $^1P_1$, $^3P_1$ and $^3P_2$ states in different computational models. Similar to extraction of the CV contribution, the effects of the CC correlation and the higher-order correlation on constants $A$ were given by the difference in results between the CC4-5 and the C$_{1s}$V models and between the MR-5 and the CC4-5 models, respectively. It can be seen that the CC correlation among electrons in the $n = 4$ core shell changes the $A$ constants by a factor of two for $^1P_1$, 14\% for $^3P_1$ and 11\% for $^3P_2$ states. On the other hand, the higher-order correlation makes contributions of 63\%, 3\% and 1\% for $^1P_1$, $^3P_1$ and $^3P_2$ states to these constants, respectively. It should be stressed that the effect of the CC correlation on the $A$ constants is opposite to the higher-order correlation effect, and thus they offset to each other partly. The similar observation was also presented in Ref.~\cite{0953-4075-41-11-115002} for the magnetic dipole hyperfine interaction constant $A$ of the ground state for Au I and in Ref.~\cite{PhysRevA.96.012514} for the $A$ constants of $3s3p~^{3,1}P_1$ states for the Al$^+$ ion. Therefore, it is essential to take into account both of them for evaluating the uncertainties of calculations. In addition, the influence of the Breit interaction and the QED corrections are about 4\%, 0.21\% and 0.33\% on the magnetic dipole hyperfine interaction constants of these three states, respectively.

\begin{table}[!ht]
\caption{\label{tab:2}Magnetic dipole hyperfine interaction constants $A$ (in MHz) for $^1P_1$, $^3P_1$ and $^3P_2$ states of $^{87}$Sr in various computational models.}
\begin{ruledtabular}
\begin{tabular}{cccc}
Models                                          &       $^1P_1$                &    $^3P_1$            &     $^3P_2$  \\ \hline
DHF                                             &        -22.77                &    -190.32            &    -166.95        \\
C$_{1s}$V                                       &         -5.10                &    -286.86            &    -236.30          \\
CC4-5                                           &        -16.12                &    -247.98            &    -210.34          \\
MR-5                                            &         -6.00                &    -256.42            &    -212.16          \\
Breit+QED                                       &         -6.24                &    -256.96            &    -212.86          \\
                                                &     \multicolumn{3}{c}{Other theories}                                      \\
Santra~\textit{et al.}~\cite{PhysRevA.69.042510}&                              &    -278               &     -231             \\
Porsev~\textit{et al.}~\cite{PhysRevA.69.042506}&                              &    -258.7             &     -211.4           \\
Boyd~\textit{et al.}~\cite{Boyd2007}            &       -15.9(5)               &                       &                      \\
Beloy~\textit{et al.}~\cite{PhysRevA.77.012512} &                              &                       &     -230.6           \\
                                                &     \multicolumn{3}{c}{Measurements}                                 \\
zu Putlitz~\textit{et al.}~\cite{ZPhys1963}     &                              &  -260.084(2)          &                      \\
Heider~\textit{et al.}~\cite{PRA161371}         &                              &                       &     -212.765(1)      \\
Kluge~\textit{et al.}~\cite{ref1}               &        -3.4(4)               &                       &                      \\
Bushaw~\textit{et al.}~\cite{Bushaw20001679}    &        -3.334(25)            &                       &                      \\
\end{tabular}
\end{ruledtabular}
\end{table}

Previous theoretical and experimental results of the magnetic dipole hyperfine interaction constants $A$ are shown in Table~\ref{tab:2} as well as the present values. It was found that our calculated $A$ constants for the $^3P_1$ and the $^3P_2$ states are in good agreement with the experimental values. Moreover, this consistency is better than theoretical results by Santra~\textit{et al.}~\cite{PhysRevA.69.042510} using an effective core potential and by Beloy~\textit{et al.}~\cite{PhysRevA.77.012512} using the configuration-interaction method (CI) coupled with many-body perturbation theory (MBPT). We also noted that our results for the magnetic dipole hyperfine interaction constants of the $^3P_1$ and the $^3P_2$ states agree well with those by Porsev~\textit{et al.}~\cite{PhysRevA.69.042506} based on similar CI+MBPT method to Beloy's. For the $^1P_1$ state, there is a relatively large difference between our calculation and measurements, due to its quite small value of the $A$ constant. Later on, we excluded this state to avoid lose of accuracy.

\begin{figure}[!ht]
 \includegraphics{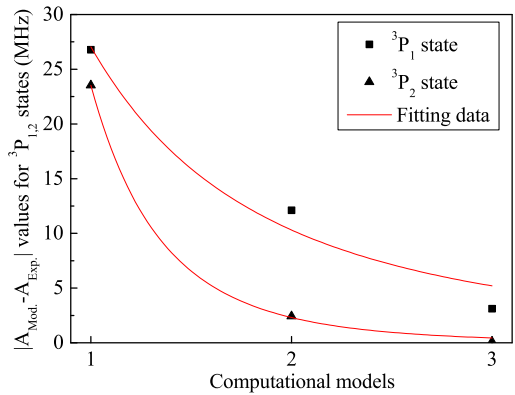}
 \caption{\label{fig:4}The absolute differences, $\left| A_{Mod.} - A_{Exp.} \right|$,  between our calculated and experimental values for the magnetic dipole hyperfine interaction constant $A$ of the $^3P_1$ and $^3P_2$ states in different computational models. The number denotes the computational model: `1' represents the C$_{1s}$V model, `2' the CC4-5 model and `3' the MR-5 model.}
\end{figure}

In order to evaluate the uncertainty of the magnetic dipole hyperfine interaction constants, we illustrate in Fig.~\ref{fig:4} the absolute value of the differences, $\left| A_{Mod.} - A_{Exp.} \right|$, between our calculated ($A_{Mod.}$) in a computational model and experimental ($A_{Exp.}$) values for the $^3P_1$ and the $^3P_2$ states. The number in this figure denotes the computational model, i.e., `1' for the C$_{1s}$V model, `2' for the CC4-5 model and `3' for the MR-5 model. As can be seen from this figure, the difference $\left| A_{Mod.} - A_{Exp.} \right|$ decreases approximately in the power of 2 as expansion of the configuration space. According to this relation, the contribution of the remained electron correlations was evaluated to be smaller than the values of $\left| A_{MR-5} - A_{Exp.} \right|$. Therefore, the uncertainties of $A$ constants for the $^3P_1$ and the $^3P_2$ states should be around 1.2\% and 0.04\%, respectively.

\subsection{Electric field gradient of $^1P_1$, $^3P_1$ and $^3P_2$ states}
In Table~\ref{tab:3}, our calculated EFGs of the $^1P_1$, $^3P_1$ and $^3P_2$ states for $^{87}$Sr are presented. We also found the offset between the CC and the higher-order correlation effects for the EFGs, similar to that in the magnetic dipole hyperfine interaction constants. As mentioned above, the $^1P_1$ states was excluded. The uncertainties of the calculated EFGs for the $^3P_1$ and the $^3P_2$ states can be evaluated based on estimation of accuracy for the magnetic dipole hyperfine interaction constants $A$, since they both have similar $r^{-3}$ dependence on the radial part of the electronic wave function~\cite{PhysRevA.71.012502, PhysRevLett.87.133003}, that is,
\begin{equation}\label{eq:10}
\Delta \mathrm{EFG}/\mathrm{EFG} \approx \Delta A/A.
\end{equation}
Therefore, the uncertainties of EFGs for these two states are about 1.2\% and 0.04\%, respectively.

\begin{table}[!ht]
\caption{\label{tab:3}Electric field gradients (in a.u.) for $^1P_1$, $^3P_1$ and $^3P_2$ states of $^{87}$Sr in various computational models.}
\begin{ruledtabular}
\begin{tabular}{cccc}
Models                  &     $^1P_1$     &    $^3P_1$           &     $^3P_2$      \\ \hline
DHF                     &       4.700     &     -0.2427          &     0.4546     \\
C$_{1s}$V               &       5.851     &     -0.5118          &     0.9658      \\
CC4-5                   &       5.572     &     -0.4075          &     0.7701       \\
MR-5                    &       5.306     &     -0.4624          &     0.8744       \\
Breit+QED               &       5.403     &     -0.4626          &     0.8749       \\
\end{tabular}
\end{ruledtabular}
\end{table}

\subsection{The nuclear electric quadrupole moment of $^{87}$Sr}
Combining our calculated EFGs of the $^3P_1$ and the $^3P_2$ states in the MR-5 model with existing experimental values of the electric quadrupole hyperfine interaction constants $B$($^3P_1$) = -35.658(6)~MHz~\cite{ZPhys1963} and $B$($^3P_2$) = 67.215(15)~MHz~\cite{PRA161371}, we obtained the nuclear electric quadrupole moment $Q$($^{87}$Sr) = 327.52~mb on average over $Q$($^3P_1$) = 328.06~mb and $Q$($^3P_2$) = 326.97~mb for the $^{87}$Sr nucleus. According to Eq.~(\ref{eq:8}), the uncertainties of EFGs lead to an error of $\Delta Q$ = 3.94~mb and 0.13~mb for the $^3P_1$ and the $^3P_2$ states, respectively. The error bars of the measured $B$ values are 0.006~MHz for the $^3P_1$ and 0.015~MHz for the $^3P_2$ states, which bring about the $\Delta Q(^3P_1)$ = 0.06~mb and $\Delta Q(^3P_2)$ = 0.07~mb. Considering all these factors, we obtained the final quadrupole moment of the $^{87}$Sr nuclei, $Q(^{87}$Sr) = 328~mb with the uncertainty of 4~mb.

For comparison, we display in Table~\ref{tab:4} other results of the electric quadrupole moments of the $^{87}$Sr nucleus. To our knowledge, zu Putlitz~\cite{ZPhys1963} reported a nuclear electric quadrupole moment of $^{87}$Sr for the first time, based on their measurement on the hyperfine structure-splitting of the $5s5p~^3P_1$ state in the $^{87}$Sr atom by optical double resonance and a simple calculation on the electric field gradient in the single-electron approximation. Later, Heider and Brink~\cite{PRA161371} extracted a $Q$ value from their experimental results of the magnetic dipole and the electric quadrupole hyperfine interaction constants of the $^3P_2$ state for $^{87}$Sr I in combination with the parametrization analysis of their results. M{\aa}rtensson-Pendrill~\cite{0953-4075-35-4-315} revised the nuclear electric quadrupole moment of $^{87}$Sr using their more accurate EFGs of the $5p~^2P_{3/2}$ state in Sr$^+$ calculated by the relativistic coupled-cluster (RCC) method. Adopting the relativistic many-body perturbation theory, Yu~\textit{et al.}~\cite{PRA70012506} obtained a $Q$ value in excellent agreement with M{\aa}rtensson-Pendrill's. Recently, however, Sahoo~\cite{PRA73062501} gave a different result of the nuclear electric quadrupole moment based on his calculated EFG of the $4d~^2D_{5/2}$ state in the Sr$^+$ ion by the RCC method. It is worth noting that the $Q$ value obtained by Sahoo was taken as a recommendation by Stone~\cite{Stone2016} and Pyykk\"o~\cite{Pekka2008, 10.1080/00268976.2018.1426131}, although the discrepancy from M{\aa}rtensson-Pendrill's and Yu~\textit{et al.} reaches about 7\%. Our result appears to confirm the $Q$ values reported by M{\aa}rtensson-Pendrill~\cite{0953-4075-35-4-315} and Yu~\textit{et al.}~\cite{PRA70012506} with respect to the consistency with each other. In addition, it should be stressed that we extracted the quadrupole moment of the $^{87}$Sr nucleus from hyperfine structures of the neutral Sr atom instead of the Sr$^+$ ion, and the EFGs were calculated in a different theoretical framework with detailed consideration of the electron correlation effects. Therefore, we would recommend our quadrupole moment $Q = 328(4)$~mb as a reference for the $^{87}$Sr nucleus.

\begin{table}[!ht]
\caption{\label{tab:4} Comparison of nuclear electric quadrupole moments $Q$ of $^{87}$Sr.}
\begin{ruledtabular}
\begin{tabular}{ccc}
                                                     &  $Q$(mb)      &      atomic systems    \\ \hline
zu Putlitz~\cite{ZPhys1963}                          &   360(3)      &      $5s5p~^3P_1$~Sr atom              \\
Heider and Brink~\cite{PRA161371}                    &   335(20)    &      $5s5p~^3P_2$~Sr atom              \\
M{\aa}rtensson-Pendrill~\cite{0953-4075-35-4-315}    &   327(24)    &      $5p~^2P_{3/2}$~Sr$^+$ ion        \\
Yu~\textit{et al.}~\cite{PRA70012506}                &   323(20)    &      $5p~^2P_{3/2}$~Sr$^+$ ion        \\
B. K. Sahoo~\cite{PRA73062501}                       &   305(2)     &      $4d~^2D_{5/2}$~Sr$^+$ ion        \\
This work                                            &   328(4)     &      $5s5p~^3P_1,~^3P_2$~Sr atom   \\
\end{tabular}
\end{ruledtabular}
\end{table}

\section{Conclusion}
The MCDHF method was employed to determine the expectation values of EFGs for the $5s5p~^3P_{1,2}$ states of the neutral Sr atom. The electron correlation effects on the EFGs, especially for the correlations related to the core shells and higher-order electron correlations, were considered systematically. We found that the contribution from the CC and the higher-order electron correlations to the EFGs are remarkable. In addition, these two effects make opposite contributions to the EFGs, and thus offset to each other partly. Therefore, it is essential to take into account
both of them for evaluating the uncertainties of calculations. Combining our calculated EFG values with measured electric quadrupole hyperfine interaction constants $B$ of these two states, we determined the nuclear electric quadrupole moment of $^{87}$Sr, $Q$ = 328(4)~mb. Our result was obtained based on the neutral atomic system rather than the Sr$^+$ ion, and the uncertainty of the present calculation on the EFGs was controlled at 1\% level. We would recommend the present $Q$ value as a reference for $^{87}$Sr.

In order to verify our result, one can extract the nuclear electric quadrupole moment of $^{87}$Sr from other atomic states if high-precision experimental values of hyperfine structures available. At present, there exist a couple of measured hyperfine structures for $5s4d~^1D_2$~\cite{Grundevik1983}, $^3D_{1,2,3}$~\cite{Bushaw1993}, $5s5d~^1D_2$~\cite{Bushaw20001679}, $5s6d~^3D_{1,2,3}$~\cite{Stellmer2014}, $5s6p~^{1,3}P_1$~\cite{Grundevik1983}, $5p4d~^1D_2$~\cite{Grundevik1983} states, but their error bars are much larger than the $Q$ value obtained in this work. Therefore, we would like to call for more accurate measurements on hyperfine interaction constants of other low-lying states for $^{87}$Sr.

\begin{acknowledgments}
This work is supported by National Natural Science Foundation of China under Grant Nos. 11874090, 91536106, 61127901, 11404025 and U1530142, West Light Foundation of The Chinese Academy of Sciences under Grant No. XAB2018B17, the Strategic Priority Research Program of the Chinese Academy of Sciences under Grant No. XDB21030700, and the Key Research Project of Frontier Science of the Chinese Academy of Sciences under Grant No. QYZDB-SSW-JSC004.
\end{acknowledgments}

\bibliography{mylib}

\end{document}